# Enhanced thermopower and low thermal conductivity in p-type polycrystalline ZrTe$_5$

M. K. Hooda and C. S. Yadav*

School of Basic Sciences, Indian Institute of Technology Mandi, Mandi-175005 (H.P.) India

Thermoelectric properties of polycrystalline *p*-type ZrTe$_5$ are reported in temperature (*T*) range *2 - 340 K*. Thermoelectric power (*S*) is positive and reaches up to *458 μV/K* at *340 K* on increasing *T*. The value of Fermi energy *16 meV*, suggests low carrier density of $\approx 9.5 \times 10^{18}$ cm$^{-3}$. A sharp anomaly in *S* data is observed at 38 K, which seems intrinsic to *p*-type ZrTe$_5$. The thermal conductivity (*κ*) value is low (*2 W/m-K* at *T = 300 K*) with major contribution from lattice part. Electrical resistivity data shows metal to semiconductor transition at *T ~ 150 K* and non-Arrhenius behavior in the semiconducting region. The figure of merit *zT* (*0.026* at *T = 300 K*) is ~ 63% higher than HfTe$_5$ (*0.016*), and better than the conventional SnTe, *p*-type PbTe and bipolar pristine ZrTe$_5$ compounds.

## INTRODUCTION

The efficient conversion of heat or thermal energy into useful electrical energy has been a key challenge for scientific community. The industrialization, decay of natural resources such as fossil fuels like oil, coal and natural gases, limitations of renewable energy sources and problem of global warming demand the highly efficient thermoelectric conversion technology[1-5]. During the past 50 years, an extensive research work is being carried out to harness the waste heat energy into useful electrical energy[5]. The dependence on primary sources of energy can be substantially reduced by finding suitable thermoelectric materials with high value of figure of merit (*zT*), which is given by the relation $zT = S^2 \sigma T/\kappa$ ;where *S*, *σ*, *κ*, and *T* are thermopower, electrical conductivity, thermal conductivity and temperature respectively[2,3]. Therefore it is desirable to have high values of *S*, *σ*, and low value of *κ* for an efficient thermoelectric material with high value of *zT*. Usually semiconductor materials are preferred over metals owing to their effective phonon scatterings, optimal carrier concentration and large phononic contribution to total *κ*. Since *S*, *σ* and *κ* are interdependent parameters, an optimum balance of these are required for achieving high value of thermoelectric efficiency (*zT*)[2-5].

Recently, a lot of attention has been paid to ZrTe$_5$, which is very well known for the peak shape resistivity anomaly at *T ~ 150 K* and bipolar conduction due to holes and electrons[4-12]. The observation of exotic physical states/properties such as 3D-topological insulator[6], Weyl semimetal[7], Zeeman splitting[8], mass acquisition of massless Dirac Fermions (electrons) [8-9], van Hove singularity near Fermi level[10], fractional Landau levels[10] has generated a lot of interest in this compound. Zhou *et al.* reported the suppression of resistivity anomaly on application of pressure and emergence of superconductivity with the $T_C$ ~ *4 K* at *14 GPa* pressure[11]. The latest report by H. Xiong *et al.*[13] has suggested 3D nature of band structure and estimated the gap of *18-29 meV* between valence and conduction band by photoemission study[13]. It is suggested that the electronic transport anomaly in ZrTe$_5$ is mediated by Lifshitz transition in the Dirac band, and strongly depends on the carrier concentration[12]. Considering the tunabliliy of carrier concentration and Fermi surface of the compound, ZrTe$_5$ can be a very promising material for thermoelectric and refrigeration applications. There have been few reports on the thermoelectric properties of the bipolar ZrTe$_5$ but the control of these properties is very difficult due to two types of carriers[4,14-18]. The electronic transport properties of ZrTe$_5$ strongly depends on various parameters such as growth conditions, impurities, doping and pressure[4,12,14-21].

We have performed electric and thermal transport measurement on *p*-type ZrTe$_5$ polycrystals. We observed a high value of thermoelectric power, leading to a substantial large *zT* value ~ *0.026* at *T = 300 K*. This value of *zT* shows that *p*-type ZrTe$_5$ is not only a better thermoelectric material than bipolar ZrTe$_5$, but other conventional materials such as PbTe and SnTe also.

## SAMPLE PREPARATION AND MEASUREMNTS

The polycrystllaline ZrTe$_5$ was synthesized by solid state reaction method, by heating of the stoichiometric elemental Zr and Te inside the evacuated quartz tube at *500 $^0$C* for *14* days, and subsequent slow cooling to *200 $^0$C*. The obtained compound was further grounded, pelletized and sintered at *500 $^0$C* for *24* hours. The single phase was confirmed by X-ray diffraction as shown in fig. 1 and lattice parameters (*a = 3.959 Å, b = 14.624 Å* and *c = 13.914 Å*) obtained from Rietveld refinement using space group *Cmcm* are close to the reported in literature[19]. The insets in fig. 1 show the crystal structure and unit cell of the compound along *ab* and *bc*-planes. The unit cell comprises large number of atoms (*24*). There are four chains of wedge shaped prisms in the unit cell, connected to each other through *Te* atoms and separated by weakly bound van der Waal gap[17]. Each pair of chains forms two dimensional structures consisting of Te atoms at the corners, forming 2D sheets with Zr atom at the center. Carriers are transported through Te sheets, from one band to other as most of the conduction electrons come from Te atoms. The electrons and holes states competition with each other with different mobilities[20] help to achieve the dominance of one type carrier in the compound, which is responsible for p-type or n-type thermopower in the compound. All the measurements ($\sigma$, $S$, and $\kappa$) were performed using Physical Properties Measurement System (PPMS) Quantum Design in $T$ range 2 K to 340 K.

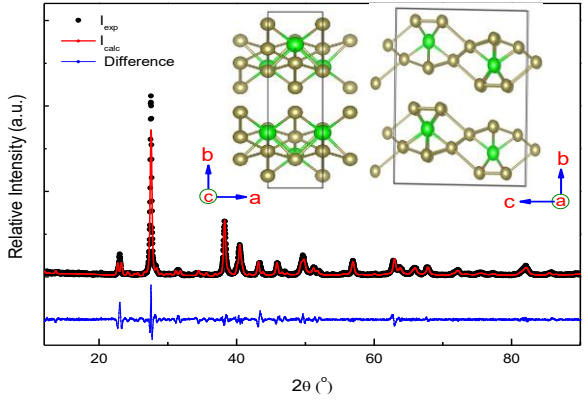

Fig. 1 Rietveld Refined X-ray diffraction pattern of ZrTe$_5$ using space group Cmcm. Insets show the unit cell drawn from the vesta software for the refined structure; green balls (Zr atoms) and yellow balls (Te atoms).

## RESULTS AND DISCUSSIONS

Figures 2 (a) shows the electrical resistivity ($\rho(T)$) measured in the $T$ range 2 - 340 K. The $\rho(T)$ exhibits metal-semiconductor transition (inset of fig.2 (a)) around 150 K. We observed a small upturn in $\rho(T)$ at $T \sim 160$ K, which is similar to that reported in the literature[4,12,14,19,22]. Shahi et al.[19] has also observed similar anomaly in ZrTe$_5$ single crystal near 150 K. The metal-semiconductor transition temperature in our ZrTe$_5$ is higher than recently reported 25 K in single crystal ZrTe$_5$ by Pariari et.al.[23] but close to reported in reference[19]. Room temperature $\rho(T)$ value of our compound 103 m$\Omega$-cm, is higher than the reported 24 m$\Omega$-cm for polycrystal ZrTe$_5$ sample of Tritt et al.[4]. However overall increase in $\rho(T)$ in switching from metal to semiconductor region from 340 K to 2 K is just 1.7 times, smaller than p-type ZrTe$_5$ single crystal (18 times from 300 K to 2K)[19]. It suggests that carrier concentrations are comparable or almost same in metallic as well as in semiconducting regions. Electronic structure calculations have also suggested ZrTe$_5$ a narrow gap semiconductor, and electrons and holes show conductivity in opposite directions[19].

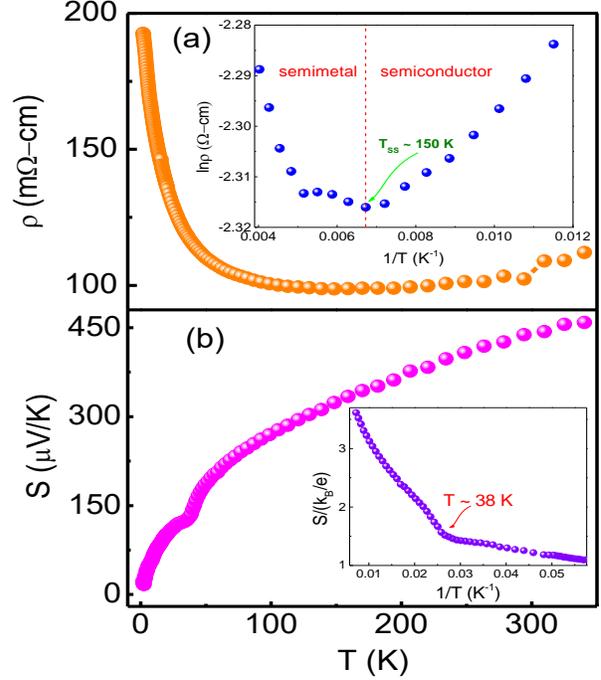

Fig. 2. (a) Electrical resistivity of ZrTe$_5$ (inset $ln\rho(T)$ vs. $1/T$ showing metal – semiconductor transition at T $\sim$ 150 K) (b) Thermopower of the compound (inset showing anomaly at 38 K).

We tried to fit $\rho(T)$ in semiconducting region for understanding transport mechanism using expression $\rho(T)=\rho_o exp(1/T^{\alpha})$ for Arrhenius ($\alpha = 1$) and Mott variable range hopping ($\alpha = 1/4$) conduction[24]. Interestingly, the plot between $Ln \rho$ and $1/T$ does not follow linear relation for $\alpha$ values of 1, 1/2, 1/3, 1/4, and 1/5, indicating the complex transport conduction mechanism in the compound. It is a possibility that transport of charge carriers takes place through the extended band states, where carriers are supplied by thermal activation of impurities states, similar to the doped semiconductors[24]. The electronic band structure of ZrTe$_5$ is known to exhibit $T$ dependence[23], and carrier concentration of transport band responsible for conduction is also $T$ dependent[19]. As tellurium gives holes to the system[19], most possible broadening happens in the density of states (DOS) of acceptor states. Exponentially decaying DOS near Fermi level, results in the broadening of accepter distribution in Fermi level[25].

The thermopower of ZrTe$_5$ (Figure 2 (b)) is positive, indicating holes as the dominant carriers in T range 2 - 340 K. The room temperature value of S in our polycrystalline compound, 458 $\mu$V/K is almost two time higher than in comparison to reported $S$ for ZrTe$_5$ and HfTe$_5$, where maximum $S$ values ranges from 75 -250 $\mu$V/K[4, 12,14-21,26]. Previous reports have shown the change in sign of $S$ from negative (n-type) to positive (p-type)[4,12,14-21]. It is to mention here that Shahi *et al.* observed anisotropic *S* which increases upon application of magnetic field[19]. It is also shown that Te vacancies (acting as *n*-type dopants) give rise to resistivity anomaly due to bipolar conduction (electrons and holes) and *S* changes its behavior from *n*-type to *p*-type near this anomaly[19]. Highly stoichiometric compound shows hole dominated *S* and semiconducting behavior in resistivity data[19]. In ref. [19], *S* values saturates in the *T* region from 60 K to 180 K but in our polycrystalline material, *S* keeps on increasing with increasing *T* up to measured 340 K. As seen from the fig 2 (b) and its inset, we have observed a kink shape anomaly at *T* ~ 38 K, which indicate the variation of electronic DOS in the compound at this temperature. Though we do not have explicit reason to account for this anomaly, similar feature has been observed in single crystal *S* data of mixed carriers (*n*-type, *p*-type) ZrTe$_5$, where application of magnetic field strengthens this anomaly[19]. However in our polycrystal ZrTe$_5$, we have only *p*-type conduction, and no change of sign of *S*. It is possibly that there is slight fluctuation in carrier concentration due to competition of two types of carriers near 38 K, which might be intrinsic to *p*-type ZrTe$_5$.

Linear fit to *S* data in the *T* range 150-325 K gives the slope value of 0.75 μV/K$^2$. The estimated Fermi energy (E$_F$) using the energy dependent relaxation time formula[27] $S = \frac{-\pi^2 k_B^2}{6e} \frac{T}{E_F}$ comes out to be ~16 meV. Such a small value of E$_F$ is expected due to low carrier concentration. Under the free electron approximation, E$_F$ gives n ≈ 9.5 × 10$^{18}$ cm$^{-3}$ which is close to the value of 5× 10$^{18}$ cm$^{-3}$ estimated for ZrTe$_5$ and HfTe$_5$ under constant scattering time approximation[19]. The value of room temperature mobility calculated using *n* and $\rho_{(T = 300 K)}$ is 6.38 cm$^2$V$^{-1}$s$^{-1}$, is very small in compare to 2970 cm$^2$V$^{-1}$s$^{-1}$ of HfTe$_5$ at 210 K[28], but consistent with the high *ρ(T)* of polycrystalline compound. The reduced mobility indicates the smaller mean free path and enhanced lattice interactions with holes.

Figure 3(a) shows *T* dependence of thermal conductivity (κ), with the room temperature value of 2.2 W/m-K, which is lower in comparison 8 W/m-K for the reported ZrTe$_5$ single crystal[29]. Our κ values are comparable to high density hot-pressed polycrystalline ZrTe$_5$, HfTe$_5$, Hf$_{0.5}$Zr$_{0.5}$Te$_5$ compounds[26]. Low value of κ could be associated with large number of atoms presents in unit cell of this pentatelluride with weak van der Waals bonding[29]. It is possibly related to the purity and stoichiometry of samples, pure sample will have less number of defects and impurities, this fact seems consistent with reduced size of maxima (5.6 W/m-K) at 19 K in comparison to single crystal (20 W/m-K) at the same *T*, because peak maxima

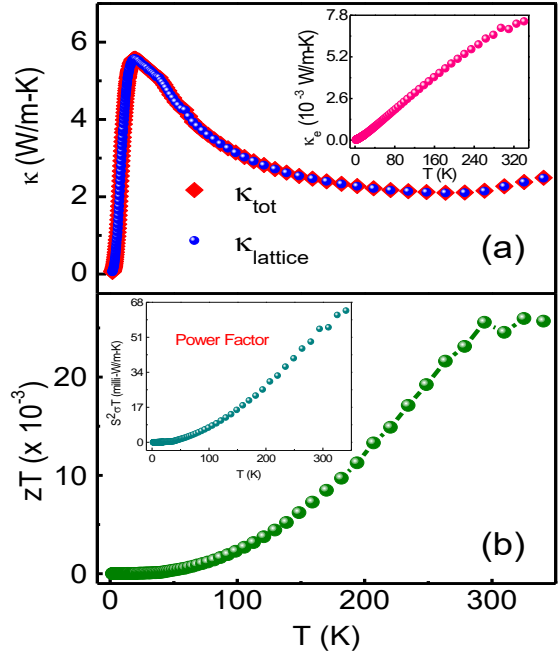

Fig. 3 (c) shows total thermal conductivity and lattice thermal conductivity (the electronic thermal conductivity is shown in inset) and (d) figure of merit (ZT) (inset shows power factor).

occurs when impurity scattering dominates over the other scattering processes[30]. For polycrystalline sample, grain boundaries contribution could be vital for κ reduction. Above 50 K, κ follows *1/T* temperature dependence, which is a signature of Umklapp[30] scattering process. The low value of κ maxima at low *T* and *1/T* dependence above 50 K suggests good quality of the sample. The electronic component κ$_e$ estimated from Wiedemann-Franz law $κ_e = LT/ρ$ where *L* is Lorentz number[30] is very low (Fig 3 (a) inset) compared to the phononic component κ$_{ph}$. The low *T* (< 6 K), κ follow usual *T$^3$* behavior due to boundary scattering[30] well consistent with very poor κ$_e$ and large κ$_{ph}$.

Figure 3 (b) shows the calculated *zT* and power factor PF (shown in inset) for the ZrTe$_5$. The *zT* value in our compound 0.026 at RT, is better in comparison to HfTe$_5$ (0.016 at RT)[26], however it is almost *4* times smaller than the high density hot pressed polycrystalline ZrTe$_5$ and Zr$_{0.5}$Hf$_{0.5}$Te$_5$[26]. Hot pressing increases the density of material and enhances σ which consequently improves the *zT* value. The *zT* value at RT is higher than for the well studied pristine SnTe, Ga and Sb, Mn doped SnTe[31,32], and almost 10 times higher than *p*-type PbTe, Ag$_2$Te[33]. In this regard, we can put ZrTe$_5$ in the category of SnTe and PbTe materials from the point of view of thermoelectric properties. Relatively lower value of *zT* against hot pressed compound is due to enhanced value of $\rho(T)$. Hot-pressing or any other similar technique possibly could further improve $\sigma$, which will consequently enhance the thermoelectric properties of the material without much affecting thermal transport properties. However the mobility for the compound is very low, this fact is supported by dominant $\kappa_{ph}$ and poor $\kappa_e$ and generally has weak $T$ dependence for semiconductors[34]. This suggests that mobility contribution to non-Arrhenius type behavior in the compound is possibly weak and carrier concentration plays a major role.

In summary, we observed non-Arrhenius $T$ dependence of electrical resistivity, devoid of variable range hopping, enhanced thermopower and reduced thermal conductivity in pristine polycrystal $p$-type ZrTe$_5$. Thermopower continues to increase with temperature up to measured 340 K. A small value of Fermi energy (16 meV) corresponding to low carrier concentration value $9.5 \times 10^{18}$ cm$^{-3}$, is optimal for good thermoelectric properties. The figure of merit shows that $p$-type ZrTe$_5$ is a better thermoelectric material than bipolar ZrTe$_5$, and other conventional materials such as PbTe and SnTe also. Suitable and optimum doping in $p$-type ZrTe$_5$ may further improve the thermoelectric properties of this material.

**ACKNOWLEDGEMENT:** We acknowledge Advanced Material Research Center (AMRC), IIT Mandi for the experimental facilities. CSY acknowledges the IIT Mandi seed grant project IITMandi/SG/ASCY/29, and DST-SERB project YSS/2015/000814 for the financial support. MKH acknowledges IIT Mandi for the HTRA fellowship.